\documentclass[pre,aps]{revtex4}
\usepackage{graphicx}

\begin{document}

\title{Multi-magnon bound states in easy-axis ferromagnetic zigzag spin chain}
\author{D.V.Dmitriev}
 \email{dmitriev@deom.chph.ras.ru}
\author{V.Ya.Krivnov}
\affiliation{Joint Institute of Chemical Physics of RAS, Kosygin
str.4, 119334, Moscow, Russia.}
\date{}

\begin{abstract}
The frustrated spin-$1/2$ chain with weakly anisotropic
ferromagnetic nearest-neighbor and antiferromagnetic
next-nearest-neighbor exchanges is studied. We focus on the
excitation spectrum and the low-temperature thermodynamics in the
ferromagnetic region of the ground state phase diagram. It is shown
that the excitation spectrum of the model is characterized by the
existence of the multi-magnon bound states. These excitations
determine the low-temperature magnetic susceptibility. The energy of
the bound magnon complexes is found using the scaling estimates of
the perturbation theory and numerical calculations. The relation of
the considered model to the edge-sharing cuprate $Li_2CuO_4$ is
discussed.
\end{abstract}

\maketitle

\section{Introduction}

The quantum spin chains with nearest-neighbor (NN) $J_1$ and
next-nearest-neighbor (NNN) interactions $J_2$ have been a subject
of numerous studies \cite{review}. The model with both
antiferromagnetic interactions $J_1$, $J_2>0$ (AF-AF model) is well
studied \cite{Haldane,Tonegawa87,Okamoto,Bursill,Majumdar,White}.
Lately, there has been considerable interest in the study of F-AF
model with the ferromagnetic NN and the antiferromagnetic NNN
interactions ($J_1<0$, $J_2>0$)
\cite{Tonegawa89,Chubukov,KO,Vekua,Lu,Nersesyan}. One of the reasons
is understanding of intriguing magnetic properties of a novel class
of quasi-one-dimensional edge-sharing copper oxides, which are
described by the F-AF model
\cite{Mizuno,Drechsler1,Drechsler2,Hase,helimagnetism1,helimagnetism2}.

The Hamiltonian of the spin-$1/2$ F-AF model is
\begin{equation}
H=J_{1}\sum_{n=1}^{N}(S_{n}^{x}S_{n+1}^{x}+S_{n}^{y}S_{n+1}^{y}+\Delta
_{1}S_{n}^{z}S_{n+1}^{z})+J_{2}
\sum_{n=1}^{N}(S_{n}^{x}S_{n+2}^{x}+S_{n}^{y}S_{n+2}^{y}+\Delta
_{2}S_{n}^{z}S_{n+2}^{z})  \label{H}
\end{equation}
where $J_1<0$ and $J_2>0$.

The isotropic case of this model ($\Delta_1=\Delta_2=1$) is
intensively studied last years \cite{Vekua,Lu,DK06,Cabra,Itoi}. The
model with the anisotropy of exchange interactions is less studied.
Though the anisotropy in real chain is weak \cite{prb134445}, it
influences on the properties of such compounds. The ground state
phase diagram of model (\ref{H}) with small anisotropy has been
studied by us in \cite{DK08}. As shown in Fig.1, the phase diagram
consists of three phases: commensurate spin-liquid gapless phase,
the incommensurate phase with spin correlations of a spiral type and
fully polarized ferromagnetic phase. Though this phase diagram is
related to the particular case of the anisotropic NN and isotropic
NNN interactions, model (\ref{H}) with both $\Delta_1\neq 1$ and
$\Delta_2\neq 1$ has the phase diagram qualitatively similar to that
shown in Fig.1 \cite{DK08}. The point $\left\vert
J_{2}/J_{1}\right\vert =1/4$ is the transition point for the
isotropic model ($\Delta_1=\Delta_2=1$), where the transition from
the ferromagnetic to the incommensurate ground state with $S^z=0$
occurs. This transition is the second order one. For the anisotropic
model the phase transition between the ferromagnetic and the
incommensurate states is of the first order type.

\begin{figure}
\includegraphics{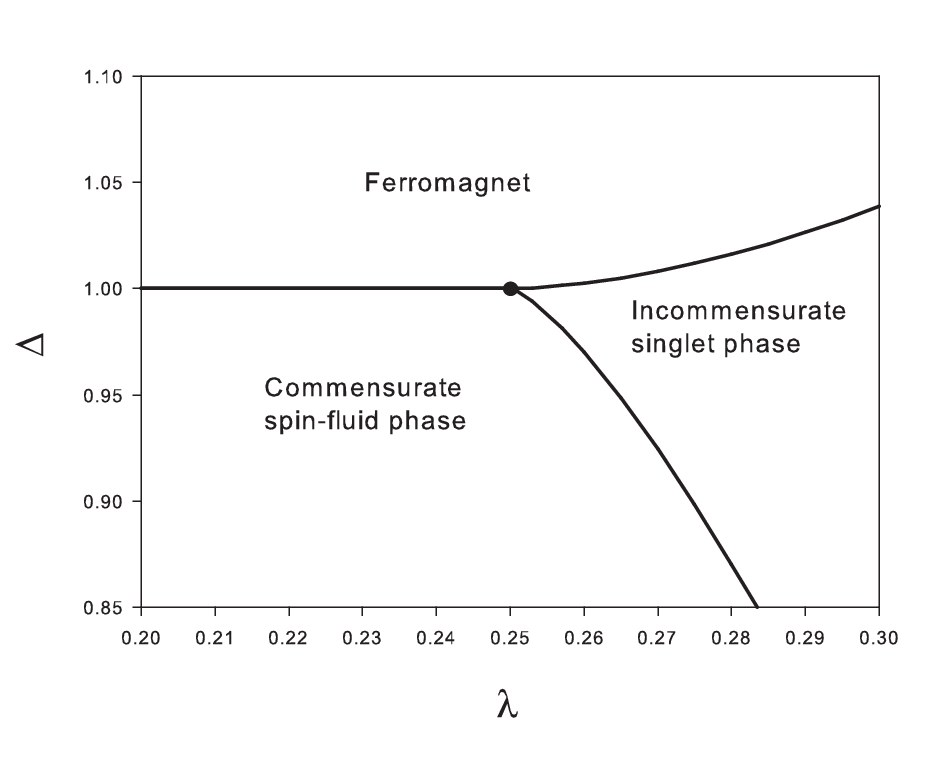}
\caption{The phase diagram of model (\ref{HH}) \cite{DK08}.}
\label{fig_1}
\end{figure}

The properties of the commensurate spin-fluid and the incommensurate
phases have been studied in \cite{DK08}. In this paper we study the
properties of the ferromagnetic (F) phase. Though the ferromagnetic
ground state is very simple, the excitation spectrum is not trivial.
The important feature of the spectrum is the existence of two types
of excitations: conventional spin waves and the multi-magnon bound
states. These excitations govern the low-temperature thermodynamics
of the model in the F phase. The additional motivation of the study
of this phase is the fact that the ferromagnetic in-chain ordering
has been observed in some edge-sharing cuprates \cite{Sapina}. This
indicates that the frustration parameter $\lambda\equiv\left\vert
J_{2}/J_{1}\right\vert$ can be less than $1/4$ in these compounds
\cite{stefan08}.

In our study we focus on the excitation spectrum and the
low-temperature thermodynamics of the F-AF anisotropic model in the
F phase. For simplicity we confine our study to the case of the
anisotropy of the NN interaction only, i.e. we consider Hamiltonian
(\ref{H}) with $\Delta_1=\Delta >1$ and $\Delta_2=1$
\begin{equation}
H=-\sum (S_{n}^{x}S_{n+1}^{x}+S_{n}^{y}S_{n+1}^{y}+\Delta
S_{n}^{z}S_{n+1}^{z}-\frac{\Delta }{4})+\lambda \sum
(\mathbf{S}_{n}\cdot \mathbf{S}_{n+2}-\frac{1}{4})  \label{HH}
\end{equation}
where we take $|J_1|$ as an energy unit and add constant shifts to
secure the energy of the fully polarized state to be zero.

We are mainly interested in the behavior of the model in the
vicinity of the transition point ($\lambda =1/4,\Delta=1$).

The paper is organized as follows. In Sec.II we represent the known
results for model (\ref{HH}) at $\lambda =0$. In Sec.III we exactly
calculate the two-magnon bound energy. In Sec.IV we perform the
scaling estimates for the energy of the multi-magnon bound states
based on the analysis of infrared divergencies in the perturbation
theory in small parameter $\alpha\equiv\Delta-1$. In Sec.V we
present results of numerical calculations of finite chains. In
Sec.VI we study the low-temperature thermodynamics of model
(\ref{HH}) and determine the region of parameters where the
multi-magnon excitations dominate. The relevance of the F-AF
anisotropic model to the copper oxide $Li_2CuO_4$ is discussed in
Sec.VII.

\section{Bound states in the ferromagnetic chain}

In this section we review the known results relevant to our study.
In the special case $\lambda =0$ model (\ref{HH}) reduces to the
ferromagnetic XXZ chain with the Ising-like anisotropy. Its ground
state is ferromagnetic with zero energy and with the gap in the
excitation spectrum. Lowest-lying excitations are bound states of
overturned spins from the fully polarized ground state (multi-magnon
bound states). The energy of $m$-magnon bound state $E_{m}(k)$ for
the chain with the periodic boundary conditions (ring) was found by
Ovchinnikov \cite{Ovchinnikov} using the Bethe Ansatz. He showed
that $E_{m}(k)$ at $N\to \infty $ is
\begin{equation}
E_{m}(k)=\frac{\sinh \left( \nu \right) \left[ \cosh (m\nu )-\cos
k\right] }{\sinh (m\nu )}  \label{Enk}
\end{equation}
where $k$ is a total momentum, $\cosh \nu =\Delta $ and
$m=1,2,\ldots $.

For $k=0$ the above equation reduces to
\begin{equation}
E_{m}\equiv E_{m}(0)=\sinh \left( \nu \right) \tanh \left(
\frac{m\nu }{2} \right)  \label{Enk0}
\end{equation}

In particular, for one- and two-magnon states it gives
\begin{eqnarray}
E_1 &=&\Delta -1  \nonumber \\
E_2 &=&\Delta -\frac{1}{\Delta }  \label{E12k0}
\end{eqnarray}
(certainly, there is no bound state for $m=1$, in this case
Eq.(\ref{Enk}) describes the one-magnon spectrum).

It follows from Eq.(\ref{Enk}) that the energy of the $m$-magnon
bound state saturates exponentially with $m$ to the value
$E_{\mathrm{s}}=\sqrt{\Delta ^{2}-1}$ and does not depend on $k$ for
$m\gg 1$, i.e. excitations become dispersionless.

Another important result for model (\ref{HH}) at $\lambda =0$ was
obtained by Alcaraz et.al. in Ref.\cite{Alkaraz}, where the
multi-magnon bound states were studied for the chain with free-end
boundary conditions (open chains). The Bethe Ansatz solution gives
the energy of the $m$-magnon bound state in the open chain in a form
\cite{Alkaraz}
\begin{equation}
E_{m,\mathrm{open}}=\sum_{i=1}^{m}(\Delta -\frac{1}{2}(\rho
_{i}-\frac{1}{\rho _{i}}))  \label{En_ro}
\end{equation}
where $\rho _{i}$ is defined by a recurrence equation
\begin{equation}
\rho_{i+1}=\frac{1}{2\Delta -\rho _{i}} \label{ro}
\end{equation}
with $\rho_1= 1/{\Delta}$.

It can be shown \cite{DKtobe} that the solution of Eq.(\ref{ro})
gives for $E_{m,\mathrm{open}}$
\begin{equation}
E_{m,\mathrm{open}}=\frac{1}{2}\sqrt{\Delta ^{2}-1}\tanh \left( m\nu
\right) \label{Enopen}
\end{equation}

The comparison of Eq.(\ref{Enopen}) with Eq.(\ref{Enk0}) leads to
the exact relation
\begin{equation}
E_{2m}=2E_{m,\mathrm{open}} \label{Enopen_ring}
\end{equation}

This relation means that the energy of the multi-magnon bound state
in the open chain at $m\gg 1$ is a half of that for the ring. This
property has been observed earlier \cite{Johnson} in numerical
calculations of model (\ref{HH}) for $\lambda =0$. Actually, the
validation of Eq.(\ref{Enopen_ring}) looks quite natural, because it
implies that the magnetic soliton of size $2m$ can be represented as
two kinks of size $m$.

As was shown in \cite{Johnson} a low-temperature thermodynamics of
the anisotropic ferromagnetic chain is determined by an effective
gap which is the lowest of two values: the gap for the one-magnon
excitations $E_1$ (spin-waves) and the gap for the multi-magnon
bound states for the \emph{open} chain (not ring)
$E_{m,\mathrm{open}}$. The comparison of Eq.(\ref{E12k0}) and
Eq.(\ref{Enopen}) shows that the effective gap for $\Delta <5/3$ is
the spin wave gap $E_1=\Delta -1$, while for $\Delta >5/3$ it is the
multi-magnon bound state energy equal to
$E_{\mathrm{kink}}=\frac{1}{2} \sqrt{\Delta ^{2}-1}$.

As will be shown below, many peculiarities of the excitation
spectrum of the anisotropic ferromagnet remain for model (\ref{HH})
with $\lambda \neq 0$.

\section{One- and two-magnon excitations}

In this Section we consider the one- and two-magnon states for the
frustrated model (\ref{HH}) with $\lambda \neq 0$. We begin with the
case of the rings with periodic boundary conditions. The energy of
the one-magnon state is
\begin{equation}
E_{1}(k)=\Delta -\cos k-\lambda (1-\cos (2k))  \label{E1k}
\end{equation}

The one-magnon spectrum has a minimum at $k=0$ for $0\leq \lambda
\leq 1/4$ and has a double-well form with two minima at $k=\pm
\arccos \left( 1/(4\lambda )\right) $ for $\lambda >1/4$. The
expansion of $E_{1}(k)$ at small $k$ ($\alpha\equiv\Delta-1$)
\begin{equation}
E_{1}(k)=\alpha +\frac{1-4\lambda }{2}k^{2}+\frac{16\lambda -1}{24}
k^{4}+\ldots  \label{E1kexpansion}
\end{equation}
shows that the behavior of the low-lying excitations on the
isotropic line $\Delta =1$ is different: in the region $0\leq
\lambda <1/4$ the low-lying excitations are described by $k^2$
spectrum
\begin{equation}
E_{1}(k)=\frac{1-4\lambda }{2}k^{2},\quad \lambda <1/4  \label{E1k2}
\end{equation}
while at the transition point $\lambda =1/4$ the one-magnon spectrum
becomes
\begin{equation}
E_{1}(k)=\frac{k^{4}}{8},\quad \lambda =1/4  \label{E1k4}
\end{equation}

As was shown in Ref.\cite{DK08} this difference plays a key role in
the change of the critical exponents near the transition point
$\lambda=1/4$.

A remarkable feature of the two-magnon spectrum is the existence of
the bound states lying below the scattering continuum. For the
isotropic F-AF model ($\Delta =1$) the two-magnon bound states at
$\lambda >1/4$ have been studied in detail earlier
\cite{Chubukov,DK06,Cabra,Kuzian,baharmuz,southern}. In particular,
it was shown that namely these states define the saturation magnetic
field in the incommensurate phase for $\lambda \gtrsim 0.367$
\cite{Kecke}.

Fortunately, the two-magnon bound state energy $E_{2}(k)$ can be
found exactly for general anisotropic case of model (\ref{HH}). The
analysis of the scattering problem of two magnons shows that for
each total momentum $k$ of the magnon pair there is one bound state.
The minimization of the energy $E_{2}(k)$ over $k$ gives the gap in
the two-magnon spectrum. In general, the dependence of $k_{\min
}(\Delta ,\lambda )$ minimizing $E_{2}(k)$ is rather complicated. We
note only that $k_{\min }\to \pi $ when $\Delta $ is increased at
fixed $\lambda $ and the dependence of the value $\Delta _{\pi
}(\lambda )$ was found in Ref.\cite{Kuzian}.

We are interested mainly in two-magnon bound states in the
ferromagnetic phase in the vicinity of the transition point $\lambda
=1/4$ and for weak anisotropy $\alpha \ll 1$. In this region the
energy $E_{2}(k)$ has a minimum at $k=0$. Therefore, hereinafter we
restrict ourselves to the magnon excitations with $k=0$. For this
case the two-magnon bound state energy $E_{2}$ is determined from
the equation:
\begin{equation}
\frac{1}{\pi }\int\limits_{0}^{\pi
}\frac{E_{1}(q)\mathrm{d}q}{E_{2}-2E_{1}(q)}=-1  \label{eqE2k0}
\end{equation}
where $E_{1}(q)$ is the spectrum of the one-magnon excitations given
by Eq.(\ref{E1k}).

Evaluating the integral in Eq.(\ref{eqE2k0}) we obtain the following
algebraic equation for $E_{2}$
\begin{equation}
\left[ (2\Delta -E_{2})^{2}-4\right] \left[ 1-2\lambda (2\Delta
-E_{2})+2\lambda \sqrt{(2\Delta -E_{2})^{2}-4}\right] =E_{2}^{2}
\label{eqE2}
\end{equation}

For $\lambda =0$ the solution of this equation reproduces the exact
result Eq.(\ref{E12k0}).

Let us represent $E_2$ in the form
\begin{equation}
E_2 = 2\alpha -\frac{\gamma^2}{4\lambda}\Theta(-\gamma) - E_b
\label{Eb}
\end{equation}
where $\Theta$ is the Heaviside function, $\gamma =1-4\lambda$ and
$E_b$ is the binding energy relative to the two-magnon scattering
continuum.

The series expansion of the binding energy in small parameter
$\alpha $ for $\lambda <1/4$ is:
\begin{equation}
E_b = \frac{\alpha^2}{\gamma }-\frac{\alpha^3}{\gamma ^{5/2}}
(1+\gamma^{1/2}-\gamma) + O(\alpha^4) \label{Ebexpansion}
\end{equation}

At the transition point $\lambda =1/4$ the solution of
Eq.(\ref{eqE2}) gives
\begin{equation}
E_{b}=\alpha ^{4/3}-\frac{2}{3}\alpha ^{5/3}+O(\alpha^2)
\label{Ebd43}
\end{equation}

Near the transition point ($\alpha\ll 1$, $|\gamma|\ll 1$)
Eq.(\ref{eqE2}) can be simplified to the cubic equation
\begin{equation}
(E_b+\gamma^2\Theta(-\gamma))^{3/2}+\gamma
(E_b+\gamma^2\Theta(-\gamma)) =\alpha^2 \label{eqEb}
\end{equation}

The solution of Eq.(\ref{eqEb}) has a scaling form depending on a
scaling variable $\kappa =\gamma /\alpha ^{2/3}$
\begin{equation}
E_b = \alpha ^{4/3}g(\kappa)  \label{Ebfx}
\end{equation}
where the function $g(\kappa)$ is a solution of equation
\begin{equation}
(g+\kappa^2\Theta(-\kappa))^{3/2}+\kappa
(g+\kappa^2\Theta(-\kappa))=1  \label{fx}
\end{equation}

In the limits $\kappa\to\infty$ ($\gamma\gg\alpha^{2/3}$) the
asymptotic of $g(\kappa)$ reproduces the leading terms of
Eq.(\ref{Ebexpansion}). In the region close to the line $\lambda
=1/4$, when $|\gamma| \ll \alpha ^{2/3}$ ($|\kappa| \ll 1$) the
expansion of $g(\kappa)=1-2\kappa/3+\ldots $ results in the
following series for the binding energy
\begin{equation}
E_b = \alpha^{4/3}-\frac23 \gamma \alpha ^{2/3} + O(\gamma^2)
\end{equation}
which contains the corrections to the leading term in
Eq.(\ref{Ebd43}).

In the region $\lambda\gtrsim\frac14$ the binding energy vanishes
on the curve $2\alpha=\gamma^2$. Near this curve according to
Eq.(\ref{eqE2}) the binding energy behaves as
\begin{equation}
E_b = \frac{(2\alpha-\gamma^2)^2}{2|\gamma|} \label{Ebexpansion2}
\end{equation}

Thus, the critical exponent characterizing a power law dependence
of the binding energy on $\alpha $ changes from $2$ at $\lambda=0$
to $4/3$ at $\lambda =1/4$. This change is due to the modification
of the behavior of the one-magnon energy at small $k$
Eqs.(\ref{E1k2}),(\ref{E1k4}).

For further study it is useful to estimate the finite-size
corrections to the two-magnon binding energy. For finite rings the
integral in equation (\ref{eqE2k0}) is replaced by a sum:
\begin{equation}
\frac{1}{N}\sum_{k}\frac{E_{1}(k)}{E_{2}-2E_{1}(k)}=-1  \label{sumk}
\end{equation}
with $k=\frac{2\pi n}{N}$ and $n=-N/2,\ldots N/2$.

The leading terms for the binding energy are governed by small $k$,
when denominators in Eq.(\ref{sumk}) are small. Therefore, we take
into account only the leading term in the expansion of $E_{1}(k)$
(Eq.(\ref{E1kexpansion})) and take infinite limits of sum in
Eq.(\ref{sumk}). So, for $\lambda <1/4$ Eq.(\ref{sumk}) reduces to
\begin{equation}
\frac{2\alpha }{N}\sum_{n=-\infty }^{\infty }\frac{1}{E_{b}+ \gamma
\left( \frac{2\pi n}{N}\right) ^{2}}=1
\end{equation}

Evaluating the sum we obtain the following equation for $E_b$
\begin{equation}
\frac{\alpha }{\sqrt{\gamma E_{b}}}\coth \left(
\frac{N\sqrt{E_{b}}}{2\sqrt{\gamma }}\right) =1
\end{equation}

At $N\gg\gamma/\alpha$ one obtains the exponentially small
finite-size correction for the leading term of the binding energy
\begin{equation}
E_{b}=\frac{\alpha ^{2}}{\gamma }(1+4e^{-N\alpha /\gamma })
\label{EbNJ0}
\end{equation}

Similar procedure for the case $\lambda =1/4$ results in the
equation
\begin{equation}
\frac{2\alpha }{N}\sum_{n=-\infty }^{\infty
}\frac{1}{E_{b}+\frac{1}{4} \left( \frac{2\pi n}{N}\right) ^{4}}=1
\end{equation}
which leads to the following equation for $E_b$
\begin{equation}
\frac{\alpha }{E_{b}^{3/4}}\frac{\sinh \left( NE_{b}^{1/4}\right)
+\sin \left( NE_{b}^{1/4}\right) }{\cosh \left( NE_{b}^{1/4}\right)
-\cos \left( NE_{b}^{1/4}\right) }=1
\end{equation}

At $N\alpha ^{1/3}\gg 1$ it gives for the binding energy
\begin{equation}
E_{b}=\alpha^{4/3} \left( 1 - \frac{8\sqrt{2}}{3}e^{-N\alpha ^{1/3}}
\sin \left( N\alpha^{1/3}+\frac{\pi}{4} \right)\right)
\label{EbNJ14}
\end{equation}

Thus, we found the exponentially small finite-size effects, which is
not a surprise for the bound states. More important is that we
identified the scaling parameters $N\alpha /\gamma $ for $\lambda
<1/4$ and $N\alpha ^{1/3}$ for $\lambda =1/4$, which will be
exploited later.

At the end of this section it is worth to make a following remark.
Let us consider the one-magnon states in the open chains. The
one-magnon problem in this case can be solved by a standard method.
The spectrum of the one-magnon excitations consists of $(N-2)$ band
states, the energies of which coincide at $N\to \infty $ with those
given by Eq.(\ref{E1k}). However, there are two degenerated bound
states localized near chain ends. Remarkably, the energy
$E_{1,\mathrm{open}}$ of these bound states is determined by
Eq.(\ref{eqE2}) with the substitution of $2E_{1,\mathrm{open}}$ for
$E_{2}$. Thus, the energy of the one-magnon bound states in the open
chain is half of the energy of the two-magnon bound state with $k=0$
in the ring. It means that relation (\ref{Enopen_ring}) remains
valid for the case $m=1$.

\section{Perturbation theory}

For more than two magnons the exact analytic solution is not
possible excluding the case $\lambda =0$. Therefore, for studying
the multi-magnon bound states we develop the perturbation theory
(PT) in small parameter $\alpha$.

At first, we inspect how the energy of the two-magnon state obtained
in the preceding section can be estimated in the framework of the
PT.

\subsection{Perturbation theory for two-magnon states}

Let us represent Hamiltonian (\ref{HH}) in a form
\begin{eqnarray}
H &=&H_{1}+\lambda V_{2}+\alpha V_{z}  \nonumber \\
H_1 &=&-\sum (\mathbf{S}_{n}\cdot \mathbf{S}_{n+1}-\frac{1}{4})
\nonumber
\\
V_2 &=&\sum (\mathbf{S}_{n}\cdot \mathbf{S}_{n+2}-\frac{1}{4})
\nonumber
\\
V_z &=&-\sum (S_{n}^{z}S_{n+1}^{z}-\frac{1}{4})  \label{PT}
\end{eqnarray}

We use the perturbation theory in small parameter $\alpha $, so that
a small perturbation $\alpha V_z$ is added to the unperturbed
Hamiltonian $H_0 = H_1 + \lambda V_2$. The ground state of $H_0$ for
$\lambda\leq 1/4$ is ferromagnetic (total spin $S=N/2$) and is
degenerate with respect to total $S^z$. The perturbation $\alpha
V_z$ splits this degeneracy. We consider the two-magnon sector,
therefore, we develop the PT to the ferromagnetic state with the
projection $S^z = N/2-2$:
\begin{equation}
\left\vert \psi _{0}\right\rangle
=\frac{1}{\sqrt{2N(N-1)}}(S^{-})^2\left\vert \uparrow \uparrow
\ldots \uparrow \right\rangle  \label{psi0}
\end{equation}
where $S^{-}=\sum S_{n}^{-}$ is the total descending operator.

In the first order of PT we have
\begin{equation}
E^{(1)}=\left\langle \psi _{0}\right\vert \alpha V_{z}\left\vert
\psi_{0}\right\rangle =2\alpha
\end{equation}

The second order reads
\begin{equation}
E^{(2)}=\alpha ^{2}\sum_{q}\frac{\left\langle \psi _{0}\right\vert
V_{z}\left\vert \psi _{q}\right\rangle \left\langle \psi
_{q}\right\vert V_{z}\left\vert \psi _{0}\right\rangle
}{E_{0}-E_{q}}  \label{E2(2)}
\end{equation}

The state $\left\vert \psi _{0}\right\rangle $ has the total
momentum $k=0$. The perturbation $V_z$ preserves the total momentum
$k$ and the projection $S^z$, but can change the total spin $S$ on
value $\pm 2$. Therefore, only two-magnon states $\left\vert \psi
_{q}\right\rangle $ with $k=0$ and $S=N/2-2$ are involved into the
PT
\begin{equation}
\left\vert \psi _{q}\right\rangle =\sum_{n,l}\varphi
_{q}(l)S_{n}^{-}S_{n+l}^{-}\left\vert \uparrow \uparrow \ldots
\uparrow \right\rangle
\end{equation}
where norm $\left\langle \psi _{q}|\psi _{q}\right\rangle =1$
requires $\varphi _{q}(l)\sim 1/N$.

The matrix elements are
\begin{equation}
\left\langle \psi _{0}\right\vert V_{z}\left\vert \psi
_{q}\right\rangle = \sqrt{\frac{2N}{N-1}}\varphi _{q}(1)\sim 1/N
\label{ME}
\end{equation}

The $N$-behavior of denominators in Eq.(\ref{E2(2)}) depends on the
value of $\lambda $. For $\lambda <1/4$, the one-magnon spectrum is
$\varepsilon_k=\gamma k^2/2$ (Eq.(\ref{E1k2})) and denominator
behaves as
\begin{equation}
E_{q}-E_{0}\sim \gamma N^{-2}  \label{EN-2}
\end{equation}

For $\lambda =1/4$, the spectrum is $\varepsilon _{k}=k^{4}/8$
(Eq.(\ref{E1k4})) and
\begin{equation}
E_{q}-E_{0}\sim N^{-4}  \label{EN-4}
\end{equation}

Thus, the estimate of the second order Eq.(\ref{E2(2)}) gives
\begin{eqnarray}
E^{(2)} &\sim &-\frac{\alpha ^{2}}{\gamma },\qquad \lambda <1/4  \nonumber \\
E^{(2)} &\sim &-\alpha ^{2}N^{2},\qquad \lambda =1/4 \label{E2}
\end{eqnarray}

From Eq.(\ref{E2}) we see that the second-order correction for the
ground-state energy converges for $\lambda<1/4$, but diverges for
$\lambda=1/4$. Therefore, we need to estimate the higher orders of
the PT. The perturbation series for the two-magnon binding energy
(Eq.(\ref{Eb})) can be written in a form:
\begin{equation}
E_{b}(\alpha )=-\left\langle \psi _{0}\right\vert \alpha
^{2}V_{z}\frac{1}{E_{0}-H_{0}}V_{z}+\alpha
^{3}V_{z}\frac{1}{E_{0}-H_{0}}V_{z}\frac{1}{E_{0}-H_{0}}V_{z}+\ldots
\left\vert \psi _{0}\right\rangle \label{E2PT}
\end{equation}

Suppose that the main contributions to the energy are given by the
low-lying excitations Eqs.(\ref{EN-2}),(\ref{EN-4}). Then the higher
orders of the perturbation series contain more dangerous
denominators and, therefore, possibly have higher powers of the
infrared divergency. Therefore, we use scaling arguments to estimate
the critical exponent for the ground-state energy. Below we will pay
attention only to the powers of the divergent terms and omit
numerical factors.

The analysis shows that the matrix elements of the perturbation
operator $V_z$ between the states $\left\vert
\psi_{q}\right\rangle $ involved in the PT behave as
\begin{equation}
\left\langle \psi _{q}\right\vert V_{z}\left\vert \psi _{q^{\prime
}}\right\rangle =N\varphi _{q}(1)\varphi _{q^{\prime }}(1)\sim 1/N
\label{ME1}
\end{equation}

Collecting the most divergent parts in all orders of the PT, we
express the correction to the binding energy as:
\begin{equation}
E_{b}=\alpha \left\langle \psi _{q}\right\vert V_{z}\left\vert \psi
_{q^{\prime }}\right\rangle \sum c_{n}x^{n}=\frac{\alpha }{N}f(x)
\label{E2nscal}
\end{equation}
where $c_{n}$ are unknown constants and
\begin{equation}
x\sim \frac{\left\langle \psi _{q}\right\vert \alpha V_{z}\left\vert
\psi _{q^{\prime }}\right\rangle }{E_{q}-E_{0}}  \label{scalpar}
\end{equation}
is a scaling parameter, which absorbs the infrared divergencies.

The estimate of scaling parameter using Eqs.(\ref{ME})-(\ref{EN-4})
gives
\begin{eqnarray}
x &=&\frac{\alpha N}{\gamma },\qquad \lambda <1/4  \nonumber \\
x &=&\alpha N^{3},\qquad \lambda =1/4
\end{eqnarray}

We note, that exactly the same scaling parameters were determined in
previous Section (Eqs.(\ref{EbNJ0}),(\ref{EbNJ14})).

In the thermodynamic limit the binding energy tends to a finite
value (independent of $N$). Therefore, the scaling function at $N\to
\infty $ behaves as
\begin{eqnarray}
f(x) &\sim &x,\qquad \lambda <1/4  \nonumber \\
f(x) &\sim &x^{1/3},\qquad \lambda =1/4
\end{eqnarray}

This gives for $E_{b}$:
\begin{eqnarray}
E_{b} &\sim &\frac{\alpha ^{2}}{\gamma },\qquad \lambda <1/4  \nonumber \\
E_{b} &\sim &\alpha ^{4/3},\qquad \lambda =1/4
\end{eqnarray}

The obtained results totally agree with those obtained in previous
Section. Thus, the scaling estimates of divergencies in the PT
correctly reproduces the scaling parameters and the leading terms in
the energy.

\subsection{Perturbation theory for $m$-magnon states}

\subsubsection{Perturbation theory at $\lambda =0$}

We start to study the multi-magnon problem from the exactly solvable
case $\lambda =0$. The lowest $m$-magnon energies in the
thermodynamic limit are given by Eq.(\ref{Enk0}). For small
anisotropy $\alpha \ll 1$ the energy $E_m$ can be written as
\begin{equation}
E_{m}(\alpha )=m\alpha f_{0}(\alpha m^{2})  \label{En_d12}
\end{equation}
with
\begin{equation}
f_{0}(x)=\sqrt{\frac{2}{x}}\tanh \left( \sqrt{\frac{x}{2}}\right)
\label{tanh}
\end{equation}

So, the expansion of Eq.(\ref{En_d12}) in small parameter $\alpha m^{2}$ is
\begin{equation}
E_{m}(\alpha )=m\alpha -\frac{1}{6}m^{3}\alpha ^{2}+\frac{1}{30}m^{5}\alpha
^{3}+\ldots  \label{EseriesJ0}
\end{equation}

However, in the opposed limit $\alpha m^{2}\gg 1$ related to large magnon
complexes, the energy converges to
\begin{equation}
E_{\mathrm{s}}(\alpha )= \sqrt{2\alpha }  \label{EmJ0}
\end{equation}

On the other hand, from the solution of two-magnon problem we know
that for $m=2$ but finite $N$, the energy depends on the scaling
parameter $x=\alpha N$ (Eq.(\ref{EbNJ0}) or Eq.(\ref{E2nscal})).
However, for $m>2$ this scaling parameter is modified and becomes
$x=\alpha mN$ \cite{DKtobe}. Thus, for the case when both $m$ and
$N$ are large but finite the energy has a scaling form of two
scaling parameters:
\begin{equation}
E_{m}(\alpha )=m\alpha f_{1}(\alpha m^{2},\alpha mN)  \label{EmNJ0}
\end{equation}
with $f_{1}(\alpha m^{2},\alpha mN)\to f_{0}(\alpha m^{2})$ at $
N\to \infty $.

\subsubsection{Perturbation theory for $0<\lambda <1/4$}

Now let us consider the case $\lambda \ll 1$. In this case PT
(\ref{PT}) contains two small parameters $\alpha $ and $\lambda $
and, consequently, two channels $V_z$ and $V_2$. Each channel can
produce infrared divergencies and is described by its own scaling
parameter \cite{DKR}. The matrix elements of the operator $V_2$ at
$\lambda \ll 1$ were found in Ref.\cite{DK08}:
\begin{equation}
\left\langle \psi _{i}\right\vert V_{2}\left\vert \psi _{j}\right\rangle
\sim N^{-2}  \label{VJexp}
\end{equation}

Such behavior of the matrix elements means that the perturbation
$V_2$ does not produce infrared divergencies and, therefore, does
not form the scaling parameter \cite{DK08}. It is natural to expect
that the behavior of the matrix elements of type (\ref{VJexp})
remains the same up to the point $ \lambda =1/4$. This assumption
results in the following expression for the lowest energy in the
sectors with small value of $S^{z}$ ($\alpha m^{2}\gg 1$) in the
region $0\leq \lambda <1/4$:
\begin{equation}
E_{\mathrm{s}}(\alpha ,\lambda )=A(\lambda )\sqrt{\alpha }  \label{Esqra}
\end{equation}
where $A(\lambda )$ is a smooth function with $A(0)=\sqrt{2}$ (see
Eq.(\ref{EmJ0})).

\subsubsection{Perturbation theory at $\lambda =1/4$}

At the transition point $\lambda =1/4$ the Hamiltonian has a form
\begin{equation}
H = H_1 + \frac14 V_2 + \alpha V_z  \label{PTJ14}
\end{equation}

In order to make the scaling estimates of the PT in $\alpha$ one
needs to know the matrix elements of the operator $V_z$.
Unfortunately, it is very difficult problem. However, we can assume
that the behavior of matrix elements with $m$ and $N$ for
$\lambda=1/4$ remains the same as for the case $\lambda=0$. Then,
the only difference between these two cases is the modification of
the spectrum presenting in denominator in Eq.(\ref{scalpar}).
According to Eqs.(\ref{EN-2}),(\ref{EN-4}) for the case $\lambda
=1/4$ both scaling parameters in Eq.(\ref{EmNJ0}) acquire additional
factor $N^2$. So, for $\lambda =1/4$ Eq.(\ref{EmNJ0}) transforms to
\begin{equation}
E_{m}(\alpha )=m\alpha f_{xy}(x,y)
\end{equation}
with $x=\alpha mN^3$ and $y=\alpha m^2N^2$.

In the thermodynamic limit, when both $x\to \infty $ and $y\to
\infty $, the scaling function $f_{xy}(x,y)$ becomes a function of
one variable (independent of $N$)
\begin{equation}
\nu =\frac{y}{x^{2/3}}=\alpha ^{1/3}m^{4/3}  \label{nu}
\end{equation}
and the $m$-magnon energy takes the form
\begin{equation}
E_{m}(\alpha )=m\alpha f_{\nu }(\nu )
\end{equation}

We expect that the expansion of the scaling function $f_{\nu }(\nu
)$ at $\nu =0$ is
\begin{equation}
f_{\nu }(\nu )=1-C_{1}\nu +C_{2}\nu ^{2}+\ldots
\end{equation}
which gives the following expansion for the energy:
\begin{equation}
E_{m}(\alpha )=m\alpha -C_{1}m^{7/3}\alpha ^{4/3}+C_{2}m^{11/3}\alpha
^{5/3}+\ldots   \label{Eseries14}
\end{equation}

So for the case $m=2$ it reduces to the obtained result for the
two-magnon binding energy Eq.(\ref{Ebd43}).

For sectors with small value of $S^{z}$ ($m\sim N/2$), the scaling
parameter $\nu \gg 1$ and the energy $E_{m}(\alpha )$ saturates to
the finite value $E_{\mathrm{s}}(\alpha )$ describing the finite
gap in the spectrum. Therefore, the asymptotic of the scaling
function $f_{\nu }(\nu )$ at $\nu \to \infty $ is $f_{\nu }(\nu
)\sim \nu ^{-3/4}$ and, finally
\begin{equation}
E_{\mathrm{s}}(\alpha )=B\alpha ^{3/4}  \label{EmJ14}
\end{equation}
where a constant $B$ will be found numerically.

\subsubsection{Perturbation theory near the transition point}

In the vicinity of the transition point the PT contains two
perturbations $\alpha V_z$ and $\frac{\gamma}{4}V_2$
\begin{equation}
H = H_1 + \frac14 V_2 + \alpha V_z - \frac{\gamma}{4} V_2
\label{PTtr}
\end{equation}

We assume that the behavior of the matrix elements of the operator
$V_2$ remains the same as in the region $\lambda <1/4$ (see
Eq.(\ref{VJexp})). At $\lambda =1/4$ according to
Eq.(\ref{scalpar}) the perturbation $\frac{\gamma}{4}V_2$ produces
the scaling parameter $z=\gamma N^2$ \cite{DK08}.

Thus, near the transition point $\lambda =1/4$ the $m$-magnon energy
can be written in a scaling form
\begin{equation}
E_{m}(\alpha ,\gamma )=m\alpha f_{xyz}(x,y,z)
\end{equation}

In the thermodynamic limit, when all parameters $x,y,z\to\infty $,
the scaling function $f_{xyz}(x,y,z)$ becomes a function of two
independent of $N$ variables: $\nu $ (Eq.(\ref{nu})) and
\begin{equation}
\mu =\frac{z\sqrt{y}}{x}=\frac{\gamma }{\sqrt{\alpha }}  \label{mu}
\end{equation}

It is worth to note that the scaling parameter $\kappa$ obtained
in the exact solution of two-magnon problem near the transition
point Eq.(\ref{Ebfx}) transforms to the parameter $\mu$
Eq.(\ref{mu}) with increase of $m$. The $m$-magnon energy in the
thermodynamic limit takes a form
\begin{equation}
E_{m}(\alpha ,\gamma )=m\alpha f_{\mu \nu }(\mu ,\nu )
\end{equation}

For the lowest states in sectors with small value of $S^z$ ($m\sim
N/2$), the scaling parameter $\nu \gg 1$ and the multi-magnon energy
becomes
\begin{equation}
E_{\mathrm{s}}(\alpha ,\gamma )=\alpha ^{3/4}f_{\mu }(\mu )  \label{Eag}
\end{equation}

The function $f_{\mu }(\mu )$ is generally unknown, however we can
determine its behavior in two limits. Close to the line $\lambda
=1/4$, when $\mu \to 0$, Eq.(\ref{Eag}) reduces to
Eq.(\ref{EmJ14}), which implies that $f_{\mu}(0)=B$. In the limit
$\mu\to\infty $ Eq.(\ref{Eag}) must reduce to Eq.(\ref{Esqra}).
This requires the asymptotic $f_{\mu }(\mu )\sim \sqrt{\mu}$ at
$\mu\to\infty $ and the energy
\begin{equation}
E_{\mathrm{s}}(\alpha ,\gamma )\sim \sqrt{\gamma \alpha },\qquad \alpha
<\gamma ^{2}
\end{equation}

This in turn means that the function $A(\lambda )$ in
Eq.(\ref{Esqra}) behaves at $\lambda \to 1/4$ as
\begin{equation}
A(\lambda )\sim \sqrt{1-4\lambda }
\end{equation}

\section{Numerical calculations}

\begin{figure*}[tbp]
\includegraphics{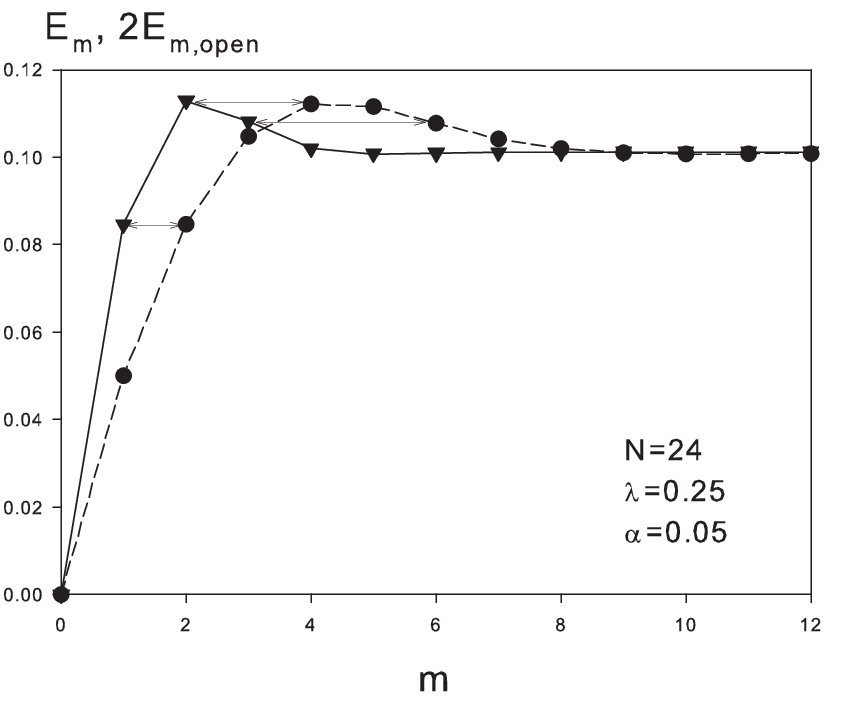}
\caption{Energies of the $m$-magnon bound states of model
(\protect\ref{HH}) for rings (circles) and open chains (triangles).
Arrows connect a few pairs of points corresponding to double
$m$-magnon energy for open chain $2E_{m,\mathrm{open}}$ and
$2m$-magnon energy for ring $E_{2m}$, validating
Eq.(\ref{Enopen_ring}).} \label{fig_2}
\end{figure*}

We have carried out exact diagonalizations of finite rings and
open chains up to 24 sites. We observed that the multi-magnon
bound states are formed when the corresponding finite-size scaling
parameters becomes large. This imposed the natural restrictions on
our calculations: $\alpha\gg 1/N^2$ for $\lambda<1/4$ and
$\alpha\gg 1/N^4$ near the transition point. We found that in all
considered cases the multi-magnon bound energies have
exponentially small finite-size corrections (for two-magnon
binding energy this fact was established analytically in Sec.II).
Therefore, we used a linear extrapolation in $\exp(-aN)$ with a
fitting parameter $a$. To check the numerical accuracy of the
extrapolation we compared the extrapolated energy for the total
$S_z=0$ ($m=N/2$) and $\lambda=0$ with Eqs.(\ref{Enk0}) and
(\ref{Enopen}) and found perfect consistency with the exact
results. Thus, for not too small $\alpha$ ($\alpha\gg 1/N^4$ near
the transition point) the use of finite chains with $N\leq 24$ is
sufficient for the extrapolation to the thermodynamic limit. We
note also that the saturation of the energy $E_m$ at $m\to N/2$
and the convergence to the thermodynamic limit for the open chains
occur noticeably faster in comparison with the rings. Really, the
kink excitation on open chain of length $N$ corresponds to the
soliton excitation on ring of length $2N$. Therefore, in our
numerical calculations we used mostly the open chains in subspace
$S^z=0$.

Firstly, we validated the important relation (\ref{Enopen_ring}).
In Section II we have shown that for $\lambda =0$ and $k=0$ the
energy of the $2m$ -magnon bound state on the ring is double of
that for the $m$-magnon bound state on the open chain. We showed
also that at $N\to\infty$ relation (\ref{Enopen_ring}) is valid
for $m=1$ and $\lambda \neq 0$. Unfortunately, we can not
rigorously prove this relation in the general case when $m>1$ and
$\lambda \neq 0$, because the analytic solution is not possible in
this case. However, we checked numerically that the relative
difference $(2E_{m, \mathrm{open}}-E_{2m})/E_{2m}$ vanishes
rapidly with $N$, so that for $N=24$ this difference is less than
$0.1\%$ for $m=1\ldots 6$ in a wide range of values of $\alpha $
and $\lambda \leq 1/4$. Therefore, we suggest that at $N\to\infty
$ the energies $2E_{m,\mathrm{open}}$ and $E_{2m}$ do coincide.
The typical dependence of the bound state energy on $m$ for rings
and open chains of length $N=24$ is demonstrated in Fig.2. As
follows from Fig.2 both energies $2E_{m,\mathrm{open}}$ and $E_m$
saturate to the same finite value when $m\gg 1$. Hence, we expect
that in the thermodynamic limit and for $m\sim N/2$ the energy of
the magnetic soliton is double the kink energy
\begin{equation}
E_{\mathrm{s}}(\alpha ,\lambda )=2E_{\mathrm{kink}}(\alpha ,\lambda)
\label{kinksol}
\end{equation}

\begin{figure*}[tbp]
\includegraphics{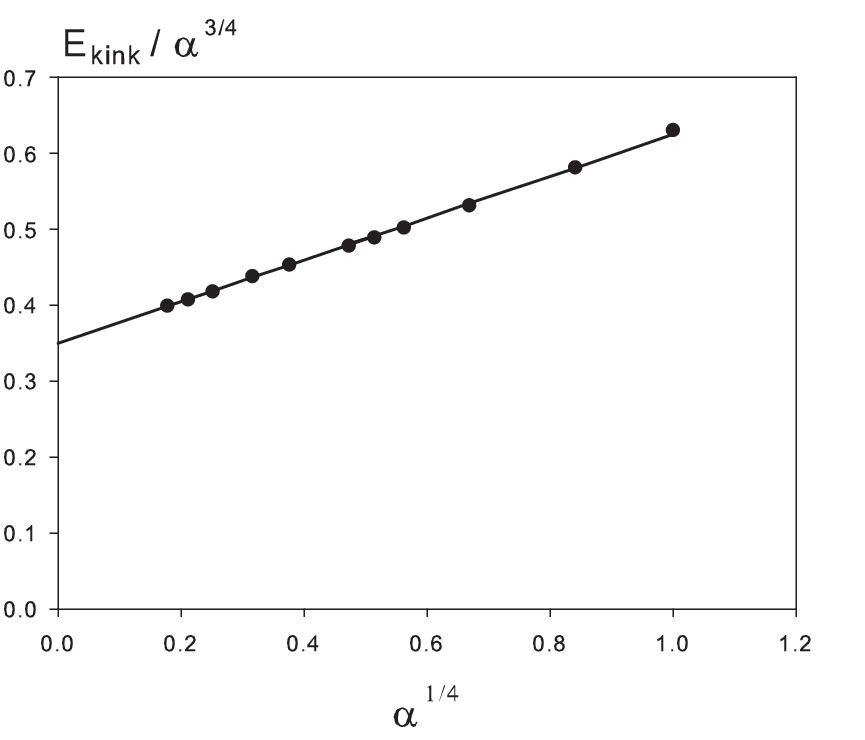}
\caption{The dependence $E_{\mathrm{kink}}/\alpha^{3/4}$ vs.
$\alpha^{1/4}$ for open chains at $\lambda =1/4$. Linear fit
corresponds to $E_{\mathrm{kink}}=0.35\alpha ^{3/4}+0.275\alpha$.}
\label{fig_3}
\end{figure*}

In Fig.3 we demonstrate the dependence of the extrapolated lowest
energy $E_{\mathrm{kink}}(\alpha)$ in the sector with $S^z=0$ for
$\lambda =1/4$. A linear fit of $E_{\mathrm{kink}}/\alpha^{3/4}$
as a function of $\alpha^{1/4}$ gives the equation
\begin{equation}
E_{\mathrm{kink}}\simeq 0.35\alpha ^{3/4}+0.275\alpha
\label{Ekink}
\end{equation}

As one can see the correction term $0.275\alpha$ gives substantial
contribution for not too small $\alpha$ and can not be neglected
(for example for $\alpha=0.1$ it gives near $30\%$ of the kink
energy). Therefore, one has to take it into account and rewrite
Eq.(\ref{Eag}) as
\begin{equation}
E_{\mathrm{kink}}=\alpha^{3/4}f_{\mu}(\mu) +\alpha g_{\mu}(\mu)
\label{Ekinkmu}
\end{equation}

The linear terms in the expansion of functions $f_{\mu}(\mu)$ and
$g_{\mu}(\mu)$ in $\mu$ correspond to the first order of PT
(\ref{PTtr}) in $\frac{\gamma}{4}V_2$ and are defined by the
behavior of the correlator
$\langle\mathbf{S}_n\cdot\mathbf{S}_{n+2}-\frac14\rangle$ at
$\lambda =1/4$. The extrapolation of numerical calculations gives
\begin{equation}
\sum_n \langle\mathbf{S}_{n}\cdot \mathbf{S}_{n+2}-\frac14\rangle
\simeq 1.92\alpha^{1/4} - 0.84\alpha^{1/2} \label{corr}
\end{equation}

This means that for $|\mu|\ll 1$ the functions $f_{\mu}$ and
$g_{\mu}$ are
\begin{eqnarray}
f_{\mu}(\mu)&\simeq &0.35 + 0.48\mu \nonumber \\
g_{\mu}(\mu)&\simeq &0.275 - 0.21\mu \label{fgmu}
\end{eqnarray}

It turns out that near the transition point it is sufficient to
take the function $g_{\mu}(\mu)$ in a form of Eq.(\ref{fgmu}).
This fact is confirmed by Fig.4, where we present the dependence
of $f_{\mu}(\mu)=(E_{\mathrm{kink}}-\alpha g_{\mu}(\mu)) /\alpha
^{3/4}$ as a function of the scaling variable $\mu =\gamma
/\sqrt{\alpha}$, calculated for different values of $\alpha$ and
$\gamma$ in the range $\alpha=0\ldots 1$ and
$\gamma=-0.2\ldots0.2$. All calculated points lie perfectly on one
curve (see Fig.4). This confirms the scaling form (\ref{Ekinkmu})
in the vicinity of the transition point. As we see in Fig.4 at
$|\mu|\ll 1$ the function $f_{\mu}(\mu)$ has the expansion
(\ref{fgmu}). In the limit $\mu\gg 1$ the numerical calculations
gives $f_{\mu}(\mu)\simeq 0.65\sqrt{\mu}$.

\begin{figure*}[tbp]
\includegraphics{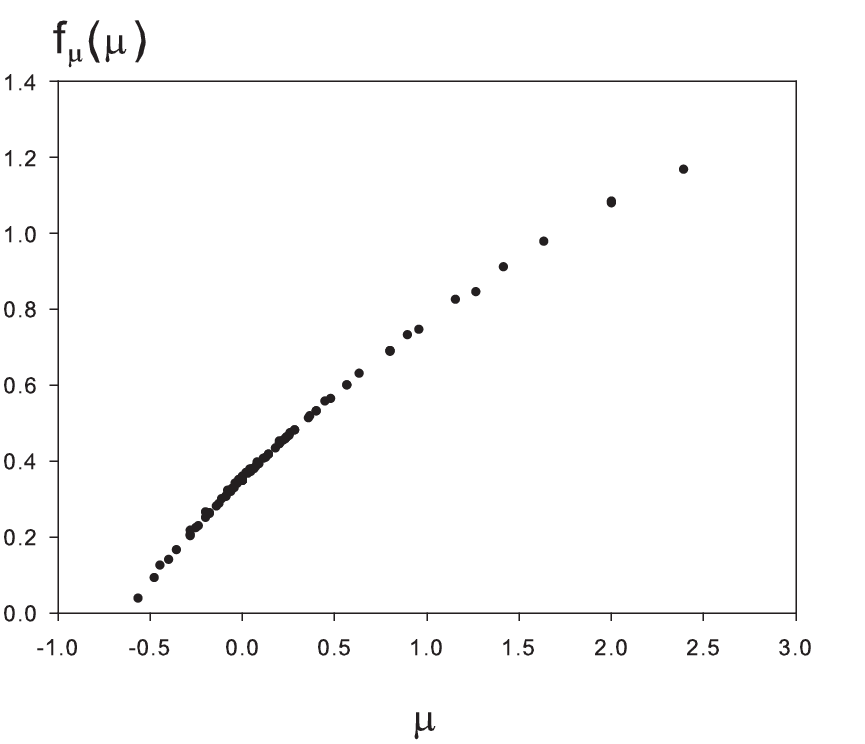}
\caption{The scaling function $f_\mu(\mu)$ in Eq.(\ref{Eag}) for
open chains. Circles correspond to different small values of
$\gamma$ and $\alpha$. Circles form perfectly one curve, justifying
the scaling dependence (\ref{Eag}).} \label{fig_4}
\end{figure*}

To end this section we list the main results obtained by numerical
calculations of finite chains for $\lambda\neq 0$. The $m$-magnon
bound state energy saturates for $m\gg 1$ for both rings and open
chains, describing the finite gap in the spectrum. The multi-magnon
bound complexes with $m\gg 1$ are very massive, which results in
flat band and quasi-degeneracy of the bound $m$-magnon excitations
for the rings over the total momentum $k$. This means that the total
degeneracy of the `soliton' energy level $E_{\mathrm{s}}$ for the
rings is proportional to $N^2$. In contrast to the rings, the
$m$-magnon bound state energy for the open chains is two-fold
degenerated for each $m$. Therefore, for the open chains the
degeneracy of the kink energy level $E_{\mathrm{kink}}$ is
proportional to $N$. The kink and soliton energies satisfy relation
(\ref{kinksol}). These properties resemble those found for
$\lambda=0$, though the energy of multi-magnon excitations is
modified for $\lambda\neq 0$. In particular, the critical exponent
depends on $\lambda$.

It may be of interest to remark one point related to the bound
magnon states in open chains. The lowest $m$-magnon state is the
state with one domain wall (kink state) and it describes the
gapped excitation above the fully polarized ground state. However,
it is possible to realize this kink as the ground state. Let us
add to Hamiltonian (\ref{HH}) the term with boundary magnetic
fields
\begin{equation}
-h(S_1^z-S_N^z)
\end{equation}

Then there is a critical magnetic field $h_c$ for which the kink
energy is zero and the kink state is degenerate with the
ferromagnetic state. It was shown in \cite{Alkaraz} that
$h_c=\frac12\sqrt{\Delta^2-1}$ for $\lambda =0$ and the critical
field $h_c$ does not depend on $m$. At $h=h_c$ the ground state is
$N$-fold degenerate. It is a result of the special symmetry of
Hamiltonian (\ref{HH}) for $\lambda =0$ with the boundary magnetic
field $h=h_c$. This is not the case for $\lambda\neq 0$. In this
case the magnetic field $h_c$ depends on $m$. For example,
$h_c(1)=\alpha^{3/4}/2^{1/4}\simeq 0.84\alpha^{3/4}$ for
$\alpha\ll 1$ and $\lambda =1/4$. The value $h_c(m)$ decreases
with $m$ and saturates to $h_c\simeq 0.35\alpha^{3/4}$ for $m\gg
1$ and $\lambda =1/4$. For $h>h_c$ the kink state with $m=N/2$
($S^z=0$) becomes the ground state.

\section{Low-temperature thermodynamics}

Results of numerical calculations show that many peculiarities of
the low-energy spectrum of model (\ref{HH}) at $\lambda \neq 0$ are
similar to those for the anisotropic ferromagnetic chain. The
thermodynamics of the latter model was studied in Ref.\cite{Johnson}
and we use the arguments of this work to treat the low-temperature
thermodynamics of model (\ref{HH}). As was shown in
Ref.\cite{Johnson} the principal contributions to the partition
function at low temperatures are given by two classes of the excited
states: spin waves and the multi-magnon bound states. Both types of
excitations are gapped and the existence of the gap implies an
exponential behavior of the thermodynamic functions at $T\to 0$.

The leading terms for the free-energy at $T\to 0$ are given by a sum
over low-lying states of both types. Certainly, the free energy must
satisfy the natural condition to be proportional to $N$. This
requires that the number of the excited states of each type must be
proportional to $N$ too. As for the spin waves the number of such
states is $\sim N$ for both rings and open chains. But it is not the
case for the multi-magnon bound states. As was shown in the
preceding section the number of these states for open chains is
$\sim N$, while for rings it is $\sim N^2$. Therefore, as was
already noticed in Sec.II, the correct contribution of the
multi-magnon excitations to the free energy is given by those for
the open chains (kinks) rather than for the rings (solitons):
\begin{equation}
F_{\mathrm{bm}}=-NT\exp \left( -\frac{E_{\mathrm{kink}}(\alpha
,\lambda )}{T} \right)  \label{Fbm}
\end{equation}
where we put Boltzmann's constant $k_B=1$.

Of course, there is no difference between open chains and rings in
the thermodynamic limit and Eq.(\ref{Fbm}) can be obtained using the
rings as well. In this case a more complicated summation of
high-lying excitations leads to the effective double reduction of
the soliton energy $E_{\mathrm{s}}$, which restores Eq.(\ref{Fbm}).

The spin-wave contribution $F_{\mathrm{sw}}$ to the free energy is
\begin{equation}
F_{\mathrm{sw}}=-\frac{NT}{\pi }\int\limits_{0}^{\pi }\exp \left(
-\frac{E_{1}(k)}{T}\right) \mathrm{d}k  \label{Fsw}
\end{equation}
where $E_1(k)$ is one-magnon energy (\ref{E1k}).

Evaluating the integral in Eq.(\ref{Fsw}) we obtain for $\lambda
<1/4$
\begin{equation}
F_{\mathrm{sw}}=-\frac{NT}{2\pi }\sqrt{\gamma }e^{(\gamma
^{2}-\alpha )/T}K_{1/4}\left( \gamma ^{2}/T\right)
\end{equation}
where $K_{1/4}$ is the modified Bessel function of the second kind.

In particular, for $T\ll\gamma^2$
\begin{equation}
F_{\mathrm{sw}}=-\frac{NT^{3/2}e^{-\alpha /T}}{\sqrt{2\pi \gamma }}
\end{equation}

For $\lambda =1/4$ ($\gamma =0$)
\begin{equation}
F_{\mathrm{sw}}=-\frac{NT^{5/4}e^{-\alpha /T}}{2^{3/4}\Gamma (3/4)}
\end{equation}

For $\lambda \gtrsim 1/4$ and $T\ll\gamma^2$
\begin{equation}
F_{\mathrm{sw}}=-\frac{NT^{3/2}e^{\gamma ^{2}/2T-\alpha
/T}}{\sqrt{\pi \left\vert \gamma \right\vert }}
\end{equation}

The dominant contribution to the low-temperature free-energy is
given by the excited states with the minimal value of the gap.
Therefore, in order to identify the prevailing type of excitations
we need to compare the kink energy $E_{\mathrm{kink}}$ with the spin
wave gap $E_{\mathrm{sw}}$, which corresponds to the minimum of
one-magnon spectrum (\ref{E1k})
\begin{eqnarray}
E_{\mathrm{sw}} &=&\alpha ,\quad \lambda <\frac14  \nonumber \\
E_{\mathrm{sw}} &=&\alpha -\frac{\gamma^2}{2},\quad \lambda \gtrsim
\frac14
\end{eqnarray}

It turns out that there are two regions in the phase plane
($\lambda,\alpha$) where the low-temperature thermodynamics is
governed by different excitations (see Fig.5). The boundary between
these two regions $\alpha_c(\lambda)$ is determined by the equation
\begin{equation}
E_{\mathrm{kink}}(\alpha_c,\lambda)=E_{\mathrm{sw}}(\alpha_c,\lambda)
\end{equation}

The calculated dependence $\alpha_c(\lambda)$ is shown in Fig.5. In
the region of the F phase below the curve $\alpha_c(\lambda)$ the
dominant excitations are spin-waves, while for
$\alpha>\alpha_c(\lambda)$ they are multi-magnon bound states. As
follows from Fig.5 the value $\alpha_c(\lambda)$ decreases when
$\lambda$ increases. In particular, $\alpha_c(0)=2/3$ and
$\alpha_c(1/4)\simeq 0.046$.

\begin{figure*}
\includegraphics{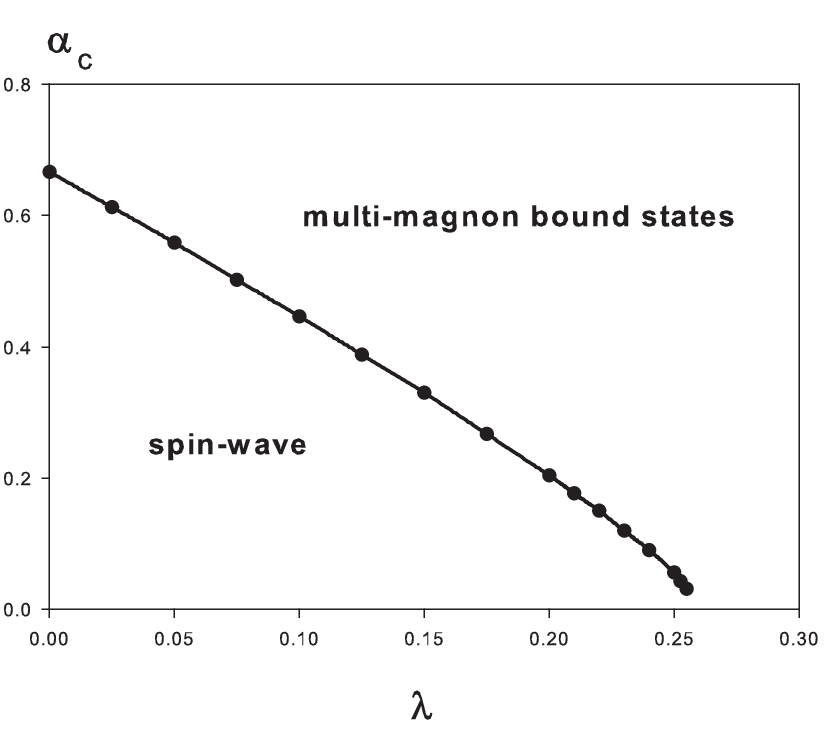}
\caption{The dependence $\alpha_c(\lambda)$. For $\alpha<\alpha_c$
the low-temperature thermodynamics is governed by spin-wave
excitations, for $\alpha>\alpha_c$ by multi-magnon bound states.}
\label{fig_5}
\end{figure*}

The specific heat in the region with the spin-wave dominance is
\begin{eqnarray}
C &=&\frac{\alpha^2 e^{-\alpha /T}}{T^{3/2}\sqrt{2\pi \gamma
}},\quad T\ll\gamma^2  \nonumber \\
C &=&\frac{\alpha^2 e^{-\alpha /T}}{T^{7/4}2^{3/4}\Gamma
(3/4)},\quad \lambda =\frac14  \nonumber \\
C &=&\frac{(\alpha -\frac12\gamma^2)^{2}e^{\gamma^2/2T-\alpha
/T}}{T^{3/2}\sqrt{\pi \left\vert \gamma \right\vert }},\quad
\lambda\geq\frac14,\quad T\ll\gamma^2
\end{eqnarray}

In the region $\alpha>\alpha_c$ where the dominant contribution to
the low-temperature thermodynamics is given by the multi-magnon
bound excitations the specific heat has an Ising-like behavior
\begin{equation}
C=\frac{E_{\mathrm{kink}}^2}{T^2}\exp\left(-\frac{E_{\mathrm{kink}}}{T}\right)
\end{equation}

To obtain the low-temperature susceptibility we add the magnetic
field $h$ along the $Z$ axis. Then the spin-wave free energy
$F_{\mathrm{sw}}$ has the same form as given by Eqs.(\ref{Fsw}) with
$\alpha $ replaced by $(\alpha +h) $ and the spin-wave contribution
to the susceptibility is
\begin{equation}
\chi _{\mathrm{sw}}=-\frac{F_{\mathrm{sw}}}{T^2}  \label{chisw}
\end{equation}

In the presence of magnetic field the energy of $m$-magnon
excitations is $E_m+mh$. The contribution to the low-temperature
susceptibility of the multi-magnon bound excitations can be obtained
using the following arguments \cite{Johnson}. At $m\gg 1$ the
magnons are tightly bound in the multi-magnon complex and the size
of the $m$-magnon bound state is $m$ with the exponentially accuracy
in $m$. Besides, these complexes are immobile because of their very
large mass. Therefore, these bound states can be considered as
domains of $m$ neighbor overturned spins in the one-dimensional
Ising model with the effective exchange constant
$E_{\mathrm{kink}}(\alpha ,\lambda )$. Using this analogy we obtain
the multi-magnon contribution to the zero-field susceptibility in
the form
\begin{equation}
\chi _{\mathrm{bm}}=\frac{1}{4T}\exp \left(
\frac{E_{\mathrm{kink}}}{T}\right)  \label{chibm}
\end{equation}

Comparing Eq.(\ref{chisw}) and Eq.(\ref{chibm}) one can easily see
that the multi-magnon contribution to the low-temperature
susceptibility is dominant at all $\alpha$ and $\lambda$ in contrast
to the case of the specific heat.

We note that equation (\ref{chibm}) determines the low-temperature
susceptibility for the anisotropic model, i.e. for $\Delta >1$ and
it is not valid in the limit $\Delta =1$, because the symmetry of
the Hamiltonian changes at $\Delta =1$. For example, the
susceptibility of the isotropic ferromagnetic Heisenberg chain
($\Delta =1$, $\lambda =0$) $\chi \sim 1/T^2$. The susceptibility
for the isotropic F-AF model can be found using the modified spin
wave approach (MSWT) \cite{MSWT}. In particular, for $\lambda =1/4$
and $\Delta =1$ this method gives \cite{DKtobe}
\begin{equation}
\chi_{\mathrm{mswt}} =\frac{1}{4(2T)^{4/3}} \label{MSWT}
\end{equation}

\section{Discussion and Summary}

Now, let us discuss a relevance of the considered model to the
copper oxide $Li_2CuO_4$. This cuprate consists of the chains formed
by the edge-sharing $CuO_4$ squares \cite{Mizuno99}. The magnetic
interaction between NN spin-$1/2$ $Cu^{2+}$ ions along the chain is
ferromagnetic while an exchange interaction between NNN $Cu$ ions is
antiferromagnetic and these chains are described by the F-AF model.
The magnetic structure of $Li_2CuO_4$ was determined by
neutron-scattering experiments \cite{Sapina}. Below the Neel
temperature $T_N\simeq 9K$ the spins of each $CuO_2$ chain has a
ferromagnetic arrangement and the arrangement between neighboring
chains is antiferromagnetic. At present the reason for the observed
ferromagnetic in-chain order is unclear. The early estimations of
the frustration parameter gave $\lambda \simeq 0.4-0.6$
\cite{Mizuno,Mizuno99}. But for such values of the frustration
parameter model (\ref{HH}) has a spiral-like ground state rather
than the ferromagnetic one \cite{Bursill,Tonegawa89}. To resolve
this discrepancy it was proposed that the observed ferromagnetic
order arises due to the specific role of the interchain interactions
\cite{Mizuno99,xiang}. However, recent estimations \cite{stefan08}
of this value based on both exact diagonalization of $Cu-O$ Hubbard
model and the DFT calculations show that the frustration parameter
is somewhat smaller than the critical value ($\lambda \simeq 0.23$).
Therefore, we can suppose that individual, non-interacting chains
have the ferromagnetic ground state. The AF long range order in
$Li_2CuO_4$ below the Neel temperature arises due to a weak
antiferromagnetic interchain interaction $J_{\perp}$.

A standard method for treating the quasi-one-dimensional systems is
the mean-field approximation for the interchain interaction. In this
approximation the Neel temperature is determined by the equation
\begin{equation}
zJ_{\perp}\chi_{1D}(T_N)=1  \label{TN}
\end{equation}
where  $\chi_{1D}$ is the susceptibility of the individual chain and
$z$ is the transverse coordination number. In $Li_2CuO_4$ each
$CuO_2$ chain is surrounded by four parallel neighboring chains and
the only non-negligible coupling $J_{\perp}$ occurs between NNN
spins on neighboring chains. Therefore, the effective coordination
number is $z=8$.

To estimate $T_N$ from Eq.(\ref{TN}) we use the susceptibility
$\chi_{1D}(T)$ given by Eq.(\ref{chibm}). According to
Ref.\cite{stefan08} the frustration parameter for $Li_2CuO_4$ is
close to the critical value and we take for the gap
$E_{\mathrm{kink}}(\alpha ,\lambda )$ its value at $\lambda =1/4$
and $\alpha \ll 1$: $E_{\mathrm{kink}}=|J_1|(0.35\alpha
^{3/4}+0.275\alpha)$. The NN in-chain interaction $J_1$ and the
interchain interaction $J_{\perp }$ were estimated in
Refs.\cite{stefan08,Graaf} as $J_1\simeq -145K$ and $J_{\perp
}\simeq 3.6K$. Using these parameters we calculated the dependence
$T_N(\alpha )$ shown in Fig.6. According to these calculations the
anisotropy in this compound is estimated as $\Delta \simeq 1.01$.
Certainly, this estimate is based on the mean-field treatment
which overestimates the transition temperature. Nevertheless, we
expect that the anisotropy of the exchange interactions in
$Li_2CuO_4$ does not exceed a few percent. This fact is also
confirmed by the estimate of the Neel temperature for the
isotropic case (see Fig.6), where we used the MSWT susceptibility
(\ref{MSWT}) in Eq.(\ref{TN}). The so obtained Neel temperature
$T_N\simeq 7.5K$ is slightly lower than the experimental value
$T_N\simeq 9K$, confirming the presence of weak anisotropy. Though
the anisotropy is very small it can essentially affect the
excitation spectrum especially when the frustration parameter is
close to the critical value $\lambda=1/4$.

\begin{figure*}
\includegraphics{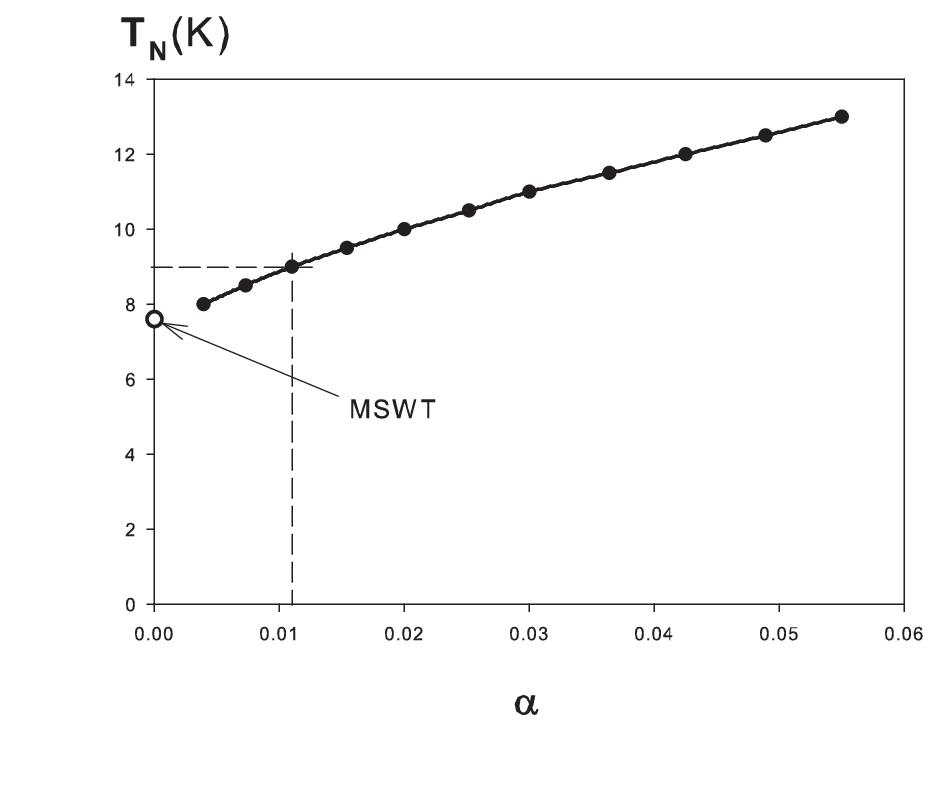}
\caption{The dependence of the Neel temperature $T_N$ on $\alpha$.
Empty circle denotes the Neel temperature for the isotropic case
calculated using the susceptibility MSWT (\ref{MSWT}).}
\label{fig_6}
\end{figure*}

In conclusion, we study the excitation spectrum of the
one-dimensional anisotropic F-AF model in a parameter range
corresponding to the ferromagnetic ground state. The remarkable
feature of the spectrum is the existence of the multi-magnon bound
states. The lowest-lying $m$-magnon excitations are
quasi-degenerated and are separated by the gap from the
ferromagnetic ground state. This gap as a function of $\alpha $ at
$\alpha \ll 1$ has a power-law behavior with the exponent
depending on the frustration parameter. It turns out that the gap
for the bound multi-magnon excitations in the rings is twice as
large of that in the open chain. The multi-magnon excitations
together with the spin waves give dominant contributions to the
low-temperature specific heat. The thermal gap characterizing the
exponential behavior of the thermodynamic functions is the
smallest value of two gaps: for one-magnon excitations and for
bound multi-magnon ones in \emph{open chains}. The comparison of
these gaps defines the regions of the dominance of one or another
type of excitations. Contrary to the specific heat the zero-field
susceptibility is always determined by the multi-magnon
excitations.

We would like to thank S.-L.Drechsler and D.Baeriswyl for valuable
comments related to this work. D.D. thanks the University of
Fribourg for kind hospitality. D.D. was supported by INTAS YS Grant
Nr. 05-- 109--4916. The numerical calculations were carried out with
use of the ALPS libraries \cite{alps}.

\end{document}